\documentclass[%
 reprint,
 amsmath,amssymb,
 aps,
 prl,
 longbibliography,
 lengthcheck,%
]{revtex4-1}

\usepackage{graphicx}% Include figure files
\usepackage{dcolumn}% Align table columns on decimal point
\usepackage{bm}% bold math
\usepackage{hyperref}% add hypertext capabilities
\begin{document}

\preprint{APS/123-QED}

\title{Collective excitations of graphene excitons being in the Bose-Einstein condensate state}% Force line breaks with \\

\author{P. A. Andreev}
\email{andreevpa@physics.msu.ru}
\affiliation{Department of General Physics, Physics Faculty, Moscow State
University, Moscow, Russian Federation.}%Lines break automatically or can be forced with \\

\date{\today}% It is always \today, today,

\begin{abstract}

Bose-Einstein condensation of the excitons in graphene is
considered. We suggested the model spinor equation for neutral
particles with short range interaction described the microscopic
graphene excitons dynamic. Using this equation we derived quantum
hydrodynamic equations for description of collective properties of
excitons in graphene, particularly for the case when excitons
being in the Bose-Einstein condensate (BEC) state. The dispersion
of collective excitations in graphene excitons BEC is studied. We
shown that frequency of collective excitations is proportional to
the square root of the wave vector module: $\omega\sim \sqrt{k}$.
\end{abstract}

\pacs{37.30.+i;  67.57.Jj;  71.35.Lk}% PACS, the Physics and Astronomy
%\keywords{}%Use showkeys class option if keyword
\maketitle

%\tableofcontents

%73.63.-b       52.27.Gr
%                    37.30.+i                 67.57.Jj Collective modes      03.75.-b Matter waves    67.57.Lm Spin dynamics
%71.35.Lk Collective effects (Bose effects, phase space filling, and excitonic phase transitions)

Bose-Einstein condensation (BEC) of ultracold atomic gases
~\cite{L.P.Pitaevskii RMP 99} and excitons in semiconductors
~\cite{Kasprzak Nat 06}-~\cite{Keeling PRL 10} \emph{and}
monolayer of carbon atoms (graphene) ~\cite{Das Sarma RMP 11},
~\cite{Peres RMP 10}, ~\cite{Goerbig RMP 12} are very interesting
and promising physical systems for present day. In this paper we
present model for excitons in graphene and study they properties
in the case one's is in the BEC state. In connection with it we
make some notice about kinematic and dynamic properties of
particles in various fundamental models.

Quantum mechanics and special relatively gives us a new way of the kinematic properties description in comparison with the classic physics. However, the methods of interaction account remains the same as in classic physics. For example, in the Schrodinger equation  the Coulomb interaction is considered as interaction of point like particles ~\cite{Landau v3}. Some times the Coulomb interaction is also used in the many-particle generalization of the Dirac's equation ~\cite{Strange book 98}.

In graphene we have deal with the electrons and holes conductivity. They properties crucially depend on fields caused by the lattice. In Ref. ~\cite{Novoselov nature 05} were found that motion of carriers in graphene might be described by Dirac's like equation. Therefore, we met a new type of kinematic properties when we study graphene carriers. In Ref.  ~\cite{Sheehy PRL
07} the Coulomb interaction were added in Hamiltonian for graphene electrons. In this paper we consider the short range interaction (SRI) between excitons in graphene and we suppose that the graphene excitons has the same kinematic properties as the graphene carriers.

Dirac's equation describe the relativistic motion of electron in external electromagnetic field. This equation solution for energy eigenvalues give us solutions with positive and negative energy separated by energy gap. Solution with the negative energies corresponds to the antiparticle states; i.e. to the positron states. When we have deal with the  semiconductors we work with the electron's and hole's. Holes are the quasi-particles corresponds to the motion "ionized" states of atoms in semiconductor. To movement of hole corresponds the movement of bound electrons in opposite direction. Electrons of conductivity moving in the space between atoms as quasi-free particles. To the hole motion correspond exchange of electron from one bound state to another between neighbor atoms. Thus, mechanisms of described two types of motion are different. This leads to the different effective mass of electrons and holes, and it's difference from mass of the free electron.

In the literature there is the analogy between behavior of relativistic electrons (the picture of electron-positron states) and electrons in semiconductors (the picture of conductivity electrons and holes). Electrons and positrons has the same mechanism of motion. Difference in the mechanisms of motion and in the effective masses of the electron's and hole's lead to the fact that using of Dirac's equation for description of the electrons with the positive and negative (hole) energy level is not always suitable for semiconductors.

Nevertheless, massless Dirac's like spinor equation has been used for the description of the conductivity electrons in graphene
\begin{equation}\label{QGGR Hamiltonian for GR}\imath\hbar\partial_{t}\psi=\Biggl(\sum_{i}\biggl(v_{F}\sigma^{\alpha}_{i}D_{i}^{\alpha}+e_{i}\varphi_{i,ext}\biggr)+\sum_{i,j\neq i}\frac{1}{2}e_{i}e_{j}G_{ij}\Biggr)\psi\end{equation}
where $\psi$ is the $N$-particle wave function which depends on 2$N$ coordinate, because graphene is the two-dimensional (2D) structure, $v_{F}$ is the Fermi velocity of conductivity electrons in graphene, $e_{i}$ electric charge, $\sigma^{\alpha}$ is the spin 1/2 Pauli matrixes, in this equation index $\alpha$ attain to $x$ and $y$, $\textbf{D}_{i}=-\imath\hbar\nabla_{i}-e_{i}/c\cdot \textbf{A}_{i, ext}$, $\varphi_{i,ext}$ and $\textbf{A}_{i, ext}$ are the scalar and vector potential of external electromagnetic field, $G_{ij}=1/r_{ij}$ is the Green function of Coulomb interaction.
The same equation might be used for description of the holes in graphene.

Electrons and holes are the charge carriers in graphene. They can form bound states - excitons. Typical velocity of electrons and holes in graphene is the Fermi velocity. The mechanisms of motion of electrons and holes in semiconductors are different, but the typical velocity of they motion in graphene is the same. Thus, we can suppose that center of mass of excitons in graphene move also with the Fermi velocity. Important difference of excitons from electrons and holes it is the fact that spin of exciton equal to integer number of Plank constant: $0$ or $\hbar$. Excition are Bose particles, whereas electrons and holes are Fermi particles.

As the atoms in quantum gases as the excitons in semiconductors
attach a lot attention in the connection with the Bose-Einstein
condensation. For the first time BEC was realized in vapors alkali
atoms in 1995. In atomic gases the BEC state reached at
hundreds of nanokelvin, whereas in exciton systems BEC state might
be realized at several kelvins ~\cite{Deng RMP 10}, ~\cite{Demenev
PRL 08},  ~\cite{Demokritov Nat 06},  ~\cite{Timofeev JAP 07}.

We suppose it is possible to realize the BEC of excitons in graphene and suggest model Dirac's like equation for the neutral particles with short range interaction for description of excitons in graphene.  This equation is an analog of the equation suggested in Ref.s ~\cite{Novoselov nature 05}, ~\cite{Sheehy PRL
07}. Here we present some notice about using of Dirac's like equation for Bose particles description. Used in Ref. ~\cite{Sheehy PRL
07} equation is linear on momentum operator as Dirac equation. Instead of scalar product of Dirac's matrixes on four momentum this equation contain the scalar product of Pauli matrixes $[\sigma_{x}, \sigma_{y}]$ on momentum operator $\textbf{p}=[p_{x}, p_{y}]$.  Thus, the model Dirac's like equation used in Ref. ~\cite{Sheehy PRL
07} might be considered as an analog of spinor Pauli equation containing another  Dirac's like kinematic. Consequently, such model equation as the Pauli equation might be used for the description as Bose as Fermi particles.

In this letter we found the general equation for the collective motion of the system of neutral spin-1 particles with SRI whose kinematic properties is analogous to graphene carriers. From this general equation we derive equation for BEC in described particles system. For derivation of equations described collective motion from the many-particle Dirac's like equation we use the method of many-particle quantum hydrodynamics (MPQHD). This method was suggested in 1999 by L. S. Kuz'menkov and S. G. Maksimov ~\cite{MaksimovTMP 1999}. Further development of the MPQHD was made in Ref.s ~\cite{MaksimovTMP 2001}-~\cite{Andreev arxiv 12 01} for various system of particles: quantum plasma ~\cite{MaksimovTMP 1999}, ~\cite{MaksimovTMP 2001}, ~\cite{Andreev arxiv MM}, particularly for spining particles ~\cite{MaksimovTMP 2001}, ~\cite{Andreev arxiv MM}, ~\cite{Mahajan PRL 11}; BEC and ultracold fermions of neutral atoms ~\cite{Andreev PRA08}; charged and neutral particles with electric dipole moment ~\cite{Andreev PRB11}, particularly electrically polarized BEC ~\cite{Andreev arxiv 12 02}; and electrons in graphene ~\cite{Andreev arxiv 12 01}. M. Marklund and G. Brodin suggested another way of derivation of QHD equations for spining particles ~\cite{Marklund PRL07}, ~\cite{Brodin NJP07}.

The QHD equations as a classic hydrodynamics are very useful for collective excitation studying ~\cite{Andreev PRB11}, ~\cite{Andreev arxiv MM}, ~\cite{Andreev arxiv 12 02}, ~\cite{Marklund PRL07}, ~\cite{Shukla RMP 11}. In this paper we consider the spectrum of collective excitations in BEC of excitons in graphene.

In this letter we do not present the detail of derivation of QHD equation for excitons BEC in graphene (GEBEC). The general scheme of QHD equations derivation is presented in Ref.s ~\cite{Andreev PRA08}, ~\cite{Andreev PRB11}, and  ~\cite{Andreev arxiv 12 01}.

We consider BEC in 2D system of particles. BEC
cannot take place in a purely 2D system at finite temperatures.
Consequently this term may be used for polaritons
under very special circumstances (see ~\cite{Amo SSaT 10} and ref.s where). Detailed discussion of the
superfluidity phenomena and BEC in 2D system is presented in Ref.
~\cite{Snoke Nat 06}. Influence of dipolar nature of excitons on
one collective properties is also studied where.

For exciton-polariton BEC description the Gross-Pitaevskii equation is used
~\cite{Cancellieri arxiv 11} as for exciton-polaritons dynamic in non-condensed states ~\cite{Adrados PRL 11}. Gross-Pitaevskii equation is usually used for studying of BEC in atomic vapors ~\cite{L.P.Pitaevskii RMP 99}. Gross-Pitaevskii equation is the non-linear Schrodinger equation. This equation has form of the Schrodinger equation for one particle in external field, but also contains the non-linear term proportional to the third degree of the wave function. Gross-Pitaevskii equation might be present in the form of two hydrodynamic equations: continuity equation and Euler equation ~\cite{L.P.Pitaevskii RMP 99}. Method of QHD allows to derive the Gross-Pitaevskii equation and it's nonlocal generalization from many-particle Schrodinger equation ~\cite{Andreev PRA08}.

We present a basic equation here for evolution description of excitons in graphene. Using this equation we derive and present below the equations for description of excitons collective motion and especially for the dynamics of GEBEC.
\begin{equation}\label{QGGR Hamiltonian}\imath\hbar\partial_{t}\psi=\Biggl(\sum_{i}\biggl(v_{F}\hat{s}^{\alpha}_{i}p_{i}^{\alpha}+V_{i,ext}\biggr)+\frac{1}{2}\sum_{i,j\neq i}U_{ij}\Biggr)\psi,\end{equation}
this equation differ from (\ref{QGGR Hamiltonian for GR}) by the form of spin matrix and the form of interaction. Equation (\ref{QGGR Hamiltonian}) contains following quantities: wave function $\psi=\psi(R,t)$, $R$ is the whole particles coordinates $R=[\textbf{r}_{1}, ..., \textbf{r}_{i}, ..., \textbf{r}_{N}]$, $\textbf{r}_{i}=[x_{i},y_{i}]$, $\hat{s}^{\alpha}_{i}$ are the spin-1 matrixes for $i$-th particle, $p_{i}^{\alpha}=-\imath\hbar\nabla$ is the momentum operator, $V_{i,ext}$ is the potential of external field, $U_{ij}$ is the SRI potential describing the interaction between excitons in graphene. We notice that at the same time some particle might interact with several particles, by means SRI potential $U_{ij}$. For this statement illustration we refer to the liquid where molecules is neutral and interacts with the several neighbor molecules. We consider spin-1 particles and spin operators are $3\times3$ matrixes
$$\begin{array}{ccc} \hat{s}_{x}=\frac{1}{\sqrt{2}}\left(\begin{array}{ccc}0&
1&
0\\
1&
0&
1\\
0&
1&
0\\
\end{array}\right),&
\hat{s}_{y}=\frac{1}{\sqrt{2}}\left(\begin{array}{ccc}0& -\imath &
0\\
\imath &
0&
-\imath \\
0&
\imath &
0\\
\end{array}\right),&
\end{array}$$
$$\hat{s}_{z}=\left(\begin{array}{ccc}1&
0&
0\\
0& 0&
0\\
0& 0&
-1\\
\end{array}\right),$$
the commutation relation for spin-1 matrixes is
\begin{equation}\label{QGGR comm rel} [\hat{s}^{\alpha}_{i},\hat{s}^{\beta}_{j}]=\imath\delta_{ij}\varepsilon^{\alpha\beta\gamma}\hat{s}^{\gamma}_{i}.\end{equation}

Equation (\ref{QGGR Hamiltonian}) describes the 2D motion of excitons, thus Hamiltonian contains operators $\hat{S}_{i}^{x}$, $\hat{S}_{i}^{y}$, $\hat{p}_{i}^{x}$ and $\hat{p}_{i}^{y}$ only, but at using commutation relation (\ref{QGGR comm rel}) during equations derivation the $S_{i}^{z}$-operator is also appearing.

Graphene is the 2D structure and electrons of graphene are located
in the plane. As we describe above in 2D case the electrons has
two coordinate $x$ and $y$, but spin of electrons can be directed
in all direction, particularly, in $z$ axes direction,
perpendicular to the graphene plane. This fact is accounted by
formula (\ref{QGGR comm rel}).

The first equation of the QHD equations set is the continuity equation
\begin{equation}\label{QGGR cont with S}\partial_{t}n(\textbf{r},t)+v_{F}\nabla\textbf{S}(\textbf{r},t)=0.\end{equation}
The definition of probability
density of conduction electron system in physical space is
\begin{equation}\label{QGGR def density}n(\textbf{r},t)=\sum_{s}\int dR\sum_{i}\delta(\textbf{r}-\textbf{r}_{i})\psi^{*}(R,t)\psi(R,t)\end{equation}
where $dR=\prod_{p=1}^{N}d\textbf{r}_{p}$. The continuity equation appears at differentiation of the concentration (\ref{QGGR def
density}) with respect to time and using equation (\ref{QGGR
Hamiltonian}) corresponding to the QHD method.

The quantity $n(\textbf{r},t)$ can be considered as 2D excitons concentration. The time evolution of
concentration  The
quantity $\textbf{S}(\textbf{r},t)$ describe the spin density of
the system of particles.
Coordinate vector $\textbf{r}$ has only two component. Consequently, equation (\ref{QGGR cont with S}) contains two component of spin density vector $\textbf{S}$, these are $S_{x}$ and $S_{y}$. Our next step in construction of the model of collective motion
is obtaining of equation of spin evolution. For this aim we
differentiate quantity $\textbf{S}(\textbf{r},t)$ with respect to time and use equation
(\ref{QGGR Hamiltonian}). Because we known the third component of spin density vector we can study the evolution of whole component of this vector. Therefore, we derive evolution equation for $\textbf{S}=[S_{x},S_{y},S_{z}]$. On this way we have equation of spin
evolution:
$$\partial_{t}S^{\alpha}(\textbf{r},t)+v_{F}\partial^{\beta}\biggl(\frac{S^{\alpha}(\textbf{r},t)S^{\beta}(\textbf{r},t)}{n(\textbf{r},t)}\biggr)$$
\begin{equation}\label{QGGR evolut of S}=-\frac{1}{\hbar}\varepsilon^{\alpha\beta\gamma}J^{\beta\gamma}_{M}(\textbf{r},t)\end{equation}%
Here new
physical quantity is appeared: $J^{\alpha\beta}_{M}(\textbf{r},t)$.
This is the tensor of spin current.
$$\partial_{t}J^{\alpha\beta}_{M}(\textbf{r},t)+\frac{v_{F}}{\hbar}\partial^{\gamma}\biggl(\frac{S^{\alpha}(\textbf{r},t)S^{\beta}(\textbf{r},t)J^{\gamma}(\textbf{r},t)}{n^{2}(\textbf{r},t)}\biggr)=$$
$$+\frac{v_{F}^{2}}{\hbar}\varepsilon^{\alpha\mu\nu}S^{\nu}(\textbf{r},t)\biggl(\frac{J^{\mu}(\textbf{r},t)J^{\beta}(\textbf{r},t)}{n^{2}(\textbf{r},t)}-\hbar^{2}\frac{\partial^{\mu}\partial^{\beta}\sqrt{n(\textbf{r},t)}}{\sqrt{n(\textbf{r},t)}}\biggr)$$
\begin{equation}\label{QGGR evol JM general}-\int d\textbf{r}' (\partial^{\beta}U(\textbf{r},\textbf{r}'))j_{2}^{\alpha}(\textbf{r},\textbf{r}',t)-v_{F}S^{\alpha}(\textbf{r},t)\partial^{\beta}V_{ext}(\textbf{r},t).\end{equation}
Equation (\ref{QGGR evol JM general}) contains interaction and a
new quantity $J^{\alpha}(\textbf{r},t)$,
$J^{\alpha}(\textbf{r},t)$ is the probability current.
$$ \partial_{t}J^{\alpha}(\textbf{r},t)+v_{F}\partial^{\beta}J^{\beta\alpha}_{M}=-v_{F}\int d\textbf{r}' (\partial^{\beta}U(\textbf{r},\textbf{r}'))\times$$
\begin{equation}\label{QGGR evol of J general}\times n_{2}^{\alpha}(\textbf{r},\textbf{r}',t)-v_{F}n(\textbf{r},t)\partial^{\alpha}V_{ext}(\textbf{r},t).\end{equation}
The equations (\ref{QGGR evol JM general}) and (\ref{QGGR evol of J general}) contains two-particle function.

These two-particle function appear in the terms contains SRI. In Ref. ~\cite{Andreev PRA08} was developed the method of calculation of the this kind quantities. Using two condition: particles interact via SRI and particles being in the BEC state, we obtain closed QHD equations set. We do not stop here on the technic of many-particles function calculation. For neutral particles being in the BEC state this technic was described in the Ref. ~\cite{Andreev PRA08}. Here we present the resulting equation found at the condition that excitons are in the BEC state.
\begin{equation}\label{QGGR evol of J final}\partial_{t}J^{\alpha}+v_{F}\partial^{\beta}J^{\beta\alpha}_{M}=v_{F}\Upsilon_{2D}n\partial^{\alpha}n-v_{F}n\partial^{\alpha}V_{ext}\end{equation}
and
$$\partial_{t}J^{\alpha\beta}_{M}+\frac{v_{F}}{\hbar}\partial^{\gamma}\biggl(\frac{S^{\alpha}S^{\beta}J^{\gamma}}{n^{2}}\biggr)=\frac{1}{2}v_{F}\Upsilon_{2D}\partial_{\beta}(S^{\alpha}n)$$
\begin{equation}\label{QGGR evol JM final}+\frac{v_{F}^{2}}{\hbar}\varepsilon^{\alpha\mu\nu}S^{\nu}\biggl(\frac{J^{\mu}J^{\beta}}{n^{2}}-\hbar^{2}\frac{\partial^{\mu}\partial^{\beta}\sqrt{n}}{\sqrt{n}}\biggr)-v_{F}S^{\alpha}\partial^{\beta}V_{ext}.\end{equation}
In these formulas we do not present the arguments of functions.
\begin{equation}\label{QGGR Upsilon} \Upsilon=\pi\int
dr(r)^{2}\frac{\partial U(r)}{\partial r},
\end{equation}
this quantity is the analog of the interaction constant in
Gross-Pitaevskii equation  ~\cite{L.P.Pitaevskii RMP 99},
~\cite{Andreev PRA08}. In the BEC theory the SRI interpret via
scattering, and $\Upsilon=-4\pi\hbar^{2}a/m,$ where $a$ is the
scattering length in first Born approximation
~\cite{L.P.Pitaevskii RMP 99}. In Ref. ~\cite{Andreev arxiv 12 02}
were presented that at derivation of QHD equations we do not
consider the scattering approximation of interaction and
connection of $\Upsilon$ and scattering length $a$ presented here
for handly comparison.

We can analyze the linear dynamics of elemental excitations in the
polarized BEC using the QHD equations (\ref{QGGR cont with S}),
(\ref{QGGR evolut of S}), (\ref{QGGR evol of J final}) and
(\ref{QGGR evol JM final}). The system is placed in
an external magnetic field $\textbf{B}_{0}=B_{0}\textbf{e}_{z}$.
The values of concentration $n_{0}$ and polarization
$\textbf{S}_{0}\parallel\textbf{B}_{0}$ for the system in an
equilibrium state are constant and uniform \emph{and} its velocity field
$J^{\alpha}(\textbf{r},t)$ and tensor
$J^{\alpha\beta}(\textbf{r},t)$ values are zero.

We consider the small perturbation of equilibrium state
$$\begin{array}{ccc}n=n_{0}+\delta n,& S^{\alpha}=S_{0}^{\alpha}+\delta S^{\alpha},& \end{array}$$
\begin{equation}\label{QGGR equlib state BEC}\begin{array}{ccc}& & J^{\alpha}=0+\delta J^{\alpha}, J^{\alpha\beta}=0+\delta J^{\alpha\beta}.\end{array}\end{equation}
Substituting these relations into system of equations (\ref{QGGR
cont with S}), (\ref{QGGR evolut of S}), (\ref{QGGR evol of J
final}) and (\ref{QGGR evol JM final}) \textit{and} neglecting
nonlinear terms, we obtain a system of linear homogeneous
equations in partial derivatives with constant coefficients.
Passing to the following representation for small perturbations
$\delta f$
$$\delta f =f(\omega, \textbf{k}) exp(-\imath\omega t+\imath \textbf{k}\textbf{r}) $$
yields the homogeneous system of algebraic equations.
The spin density strength is assumed to have a nonzero value.
Expressing all the quantities entering the system of equations in
terms of the spin density, we come to the equation
\begin{equation}\label{QGGR lin equation}\Lambda^{\alpha\beta}(\omega, \textbf{k}) S^{\beta}(\omega, \textbf{k})=0,\end{equation}
Excitations exist in the case when matrix equation (\ref{QGGR lin
equation}) has nontrivial solutions. The condition of nontrivial
solution existence for homogeneous linear algebraical equation set
is the determinant of this equation set must be equal to zero:
\begin{equation}\label{QGGR determinant}det\parallel\Lambda^{\alpha\beta}(\omega, \textbf{k})\parallel=0.\end{equation}
Solving this equation with respect to $\omega^{2}$ we obtain a
following results.

Equation (\ref{QGGR determinant}) give us only one wave solution,
dispersion of this wave is obtained in the form:
\begin{equation}\label{QGGR dispertion spectrum} \omega^{2}=\frac{v_{F}n_{0}\mid\Upsilon\mid k}{\hbar\sqrt{2}},\end{equation}
where $k^{2}=k_{x}^{2}+k_{y}^{2}$.

Solution (\ref{QGGR dispertion spectrum}) does not depend on the
sign of SRI, and contain module of the SRI constant $\Upsilon$.

For comparison we present here the dispersion of the Bogoliubov's mode which is the eigenwave in usual BEC:
\begin{equation}\label{QGGR Bogol} \omega^{2}=\frac{\hbar^{2}}{4m^{2}}k^{4}+g\frac{ n_{0}}{m}k^{2}, \end{equation}
in the long wavelength limit from formula (\ref{QGGR Bogol}) we
have $\omega^{2}=gn_{0}k^{2}/m$. The quantity $g$ is the
interaction constant in Gross-Pitaevskii equation
~\cite{L.P.Pitaevskii RMP 99} and $g=-\Upsilon$. This solution
exist for repulsive interaction ($g>0$, $a>0$) only.

Thus, instead of sound like spectrum $\omega\sim k$ for BEC in atoms vapors for the GEBEC we have $\omega\sim \sqrt{k}$.

In connection with the studying developments of the BEC in
excitons system in semiconducturs and wide using of graphene in
semiconductor geterostructures we suggest the model for
description of GEBEC collective properties. One of the fundamental
properties of many-particle systems is the spectrum of collective
excitation. Thus, we considered the collective excitation spectrum
for GEBEC.

Using the idea  that excitons in graphene has an analogous
kinematic properties as the graphene carriers we formulated
many-particle Dirac's like equation (or linear on momentum pauli
like equation) for spin-1 neutral particles with SRI, describing
graphene excitons dynamics. Starting from this equation, using
method of many-particle QHD we derived the system of QHD
equations: particles number balance equation (continuity
equation), spin balance equation, current evolution equation
(Euler equation) and equation evolution of spin current. We
considered graphene excitons being in the BEC state and studied
the dispersion of collective excitations where. We found that
derived model leads to existence of one type of collective
excitations in GEBEC. Dispersion dependence of these excitations
differ from Bogoliubov's spectrum, which suitable for BEC of
atomic vapors or liquid helium. In the case GEBEC the frequency of
collective excitations is proportional to $\sqrt{k}$. Thus,
obtained dependence differ from sound wave and long wave length
limit of Bogoliubov's spectrum.

The author thanks Professor L. S. Kuz'menkov for fruitful
discussions.

%\nocite{*}

\end{document}